\documentclass[aps,floatfix,twocolumn,prb,superscriptaddress]{revtex4-2}
\usepackage{xcolor} 
\usepackage{graphicx}
\usepackage[hidelinks]{hyperref}
\usepackage{amssymb}
\usepackage[utf8]{inputenc}
\usepackage{amsmath}
\usepackage{todonotes}

\newcommand{\op}[1]{\hat{\mathrm{#1}}}
\renewcommand{\vec}[1]{\boldsymbol{#1}}

\newcommand{\change}[1]{\textcolor{black}{#1}}

\renewcommand{\section}[1]{}

\definecolor{vastkust}{RGB}{0, 48, 80} 
\hypersetup{
  colorlinks,
  citecolor=vastkust,
  linkcolor=vastkust,
  urlcolor=vastkust}

\begin{document}
\title{Correlation Driven Magnetic Frustration and Insulating Behavior of TiF$_3$}

\author{Gayanath W.~Fernando}
\email{gayanath.fernando@uconn.edu}
\affiliation{Department of Physics, University of Connecticut, Storrs, Connecticut, 06269 USA}

\author{Donal~Sheets }
\affiliation{Department of Physics, University of Connecticut, Storrs, Connecticut, 06269 USA}

\author{Jason~Hancock }
\affiliation{Department of Physics, University of Connecticut, Storrs, Connecticut, 06269 USA}

\author{Arthur~Ernst}
\affiliation{Max Planck Institute of Microstructure Physics, Weinberg 2, D-06120 Halle, Germany}
\affiliation{Institute for Theoretical Physics, Johannes Kepler University, Altenberger Strasse 69, 4040 Linz, Austria}

\author{R.~Matthias Geilhufe}
\email{matthias.geilhufe@chalmers.se}
\affiliation{Department of Physics, Chalmers University of Technology, 412 96 G\"{o}teborg, Sweden}

\date{\today}
\begin{abstract}
We investigate the halide perovskite TiF$_3$, renowned for its intricate interplay between structure, electronic correlations, magnetism, and thermal expansion. Despite its simple structure, understanding its low-temperature magnetic behavior has been a challenge. \change{Previous theories proposed antiferromagnetic ordering. In contrast, experimental signatures for an ordered magnetic state are absent down to 10~K}. Our current study has successfully reevaluated the theoretical modeling of TiF$_3$, unveiling
the significance of strong electronic correlations as the key driver for its insulating behavior and magnetic frustration.
In addition, our frequency-dependent optical reflectivity measurements exhibit clear signs of an insulating state.
Analysis of the calculated magnetic data gives an antiferromagnetic exchange coupling with a net Weiss temperature of order 25~K as well as a magnetic response consistent with a $S$=1/2 local moment per Ti$^{3+}$. Yet, the system shows no susceptibility peak at this temperature scale and appears free of long-range antiferromagnetic order down to 1~K. Extending ab initio modeling of the material to larger unit cells shows a tendency for relaxing into a non-collinear magnetic ordering, with a shallow energy landscape between several magnetic ground states, promoting the status of this simple, nearly cubic perovskite structured material as a candidate spin liquid. 
\end{abstract}
\maketitle

\section{Introduction}

Perovskite materials are known to exhibit fascinating 
physical properties such as colossal magnetoresistance, high-temperature superconductivity and negative thermal expansion.  Fluoride perovskites,
with the typical formula unit MF$_3$, only require a trivalent M site in contrast to a typical AMO$_3$ for oxide perovskites. Transition metal trifluoride perovskites are less well studied compared to their oxide counterparts, and known to display marked differences such as the strong negative thermal expansion (NTE) in ScF$_3$. Doping of such MF$_3$ perovksites has been the focus of the experimental work of Morelock et al.~\cite{morelock2014} which provides a fairly comprehensive structural study of the material class Sc$_{1-x}$Ti$_{x}$F$_3$. Such materials have both fundamental and practical value. When designing composite materials, controlling and tuning their thermal properties is highly desirable to avoid thermal fracture/fatigue and related issues.

Most of the materials mentioned above also show a cubic to rhombohedral structural transition as temperature is lowered. Above 350-370 K, the trifluoride TiF$_3$ has the cubic perovskite structure AMX$_3$ with no cation A present. Each Ti is at the center of corner sharing octahedrons of fluorine atoms. Perturbations of the strong NTE state using metallic substitution A on the trivalent Sc$^{+3}$ site in Sc$_{1-x}$Al$_{x}$F$_3$ has revealed the NTE behavior to be present at elevated temperatures for A = Al, Y, Ti, Fe over a significant range of compositions. The opportunity to compositionally introduce electronic degrees of freedom in a system with a structurally critical ground state is an intriguing prospect, particularly in light of the high interest in low-carrier superconductivity in perovskites SrTiO$_3$ and KTaO$_3$. There are reasons to believe that the structural phase transition from cubic to rhombohedral is not directly tied to magnetic properties of the metal atom A since even AlF$_3$ exhibits it. In fact, induced dipole-dipole interactions of the F ions are believed to drive this transition~\cite{allen}.   

In our opinion, the magnetic and insulating properties of TiF$_3$ have not been well understood amidst several controversial claims. The extra valence $d$-electron that Ti carries compared to Sc (\cite{morelock2014, sowa1998}) gives rise to unusual electronic and magnetic properties.

Several theoretical studies have been reported in the past, discussing the magnetic ground state of TiF$_3$. Perebeinos and Vogt employing the FLAPW method \cite{Perebeinos2004} showed that without correlation effects, TiF$_3$ would be a ferromagnetic half-metal. If the Hubbard interaction strength $U$ exceeds a value of $4 J_H$, with $J_H$ being the Hund-coupling, the material undergoes a transition into a Mott insulator. Similar behavior was later confirmed, e.g., by Mattsson and Paulus \cite{mattsson2019density}, using a slightly revised computational approach. In contrast, Qin \textit{et al.} \cite{Qin2022} claimed that in metal trifluorides, itinerant electrons play a central role in controlling thermal expansion while ignoring strong electron correlation effects mentioned in the FLAPW work above~\cite{Perebeinos2004}. 
Even though theory work hints towards an antiferromagnetic state, it has been mentioned that low temperature neutron scattering experiments have not revealed any such transition \cite{Perebeinos2004}. More recent experiments by Sheets \textit{et al.} \cite{sheets2023} show clear indications for some kind of magnetic transition below 10~K, hinting towards a more complex magnetic order or magnetic frustration.  

In the following, we discuss the option of magnetic frustration by extending the theoretical modeling of TiF$_3$ towards non-collinear magnetic configurations. We show that such a solution gives rise to an insulating ground state in agreement with experiment. In addition, it explains the absence of a clear antiferromagnetic transition. 

\section{Phase Transitions - phenomenological theory}
\begin{figure}[b!]
    \centering
    \includegraphics[width=0.49\textwidth]{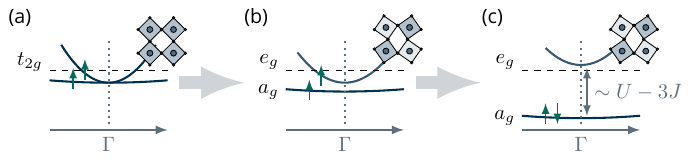}
    \caption{Illustration of symmetry breaking order in TIF$_3$. (a) in the normal state and in the absence of interactions, TIF$_3$ would behave as a ferromagnetic metal. (b) due to the partial occupation of the $t_{2g}$ band the total energy of the system can be lowered by a Jahn-Teller-like symmetry breaking, lifting the band degeneracy. (c) strong Coulomb repulsion due to flat bands impose a significant band splitting between the $e_g$ and $a_g$ sates, favoring an insulating state with antiferromagnetic or frustrated magnetic order.  }
    \label{fig:phases}
\end{figure}

At high temperatures, TiF$_3$ is cubic (space group Pm$\overline{3}$m). Each Ti ion contains an unpaired electron occupying a $t_{2g}$ orbital. Due to the partial occupation of the $t_{2g}$ band, the system can lower its free energy by a rhombohedral distortion and a resulting symmetry breaking into the space group $R\overline{3}c$ taking place at temperatures below 350~K. This lowering of symmetry lifts the degeneracy of the $t_{2g}$ band, imposing a split into two bands with $e_g$ and $a_g$ character which corresponds to the point group $D_{3d}$ (Figures \ref{fig:phases} (a) and (b)). However, this splitting is very weak $\approx 6$~meV as opposed to the total band width of $\approx 2$~eV and an accompanying band narrowing effect of $\approx 18$~meV \cite{Perebeinos2004}.  The corresponding basis functions transform as \cite{Perebeinos2004,gtpack1,gtpack2}
\begin{equation}
    \left|a_g\right> = \frac{\left|xy\right>+\left|yz\right>+\left|zx\right>}{\sqrt{3}},
    \label{orbitals1}
\end{equation}
and
\begin{equation}
   \left|e_g;1\right> = \frac{\left|yz\right>-\left|zx\right>}{\sqrt{2}}, \quad \left|e_g;2\right> = \frac{2\left|xy\right>-\left|yz\right>-\left|zx\right>}{\sqrt{6}}.  
   \label{orbital2}
\end{equation}

To first order, the transition is described as a tilting of corner sharing octahedrons surrounding each metal atom. TiF$_3$ has been observed to undergo the phase transition from rhombohedral to cubic with noticeable positive thermal expansion below the transition and
negligible thermal expansion above it. Fluorine is bonded in a linear geometry to two equivalent Ti atoms with a bond length of 1.97 {\AA} in the cubic phase. The tilting angle has been observed to vary between the cubic value (i.e., zero degrees) to a maximum of about sixteen degrees (continuously) in the rhombohedral phase as temperature is lowered, and could play a role in controlling a possible magnetic super-exchange interaction between Ti atoms. Other Ti doped Sc-trifluorides (Sc$_{1-x}$Ti$_x$F$_3$) also display similar phase transitions occurring at lower temperatures\cite{morelock2014}. However, note that there are trifluorides such as AlF$_3$ with no $d$-electrons where similar tilting of octahedrons has been observed and discussed in terms of electrostatic dipole-dipole interactions due to displaced Fluorine ions~\cite{allen}. While it is likely that such dipole-dipole interactions are present in TiF$_3$,
we believe that the resulting gap and magnetism is directly tied to
electronic correlations.

\begin{figure}
    \centering
    \includegraphics[width=0.49\textwidth]{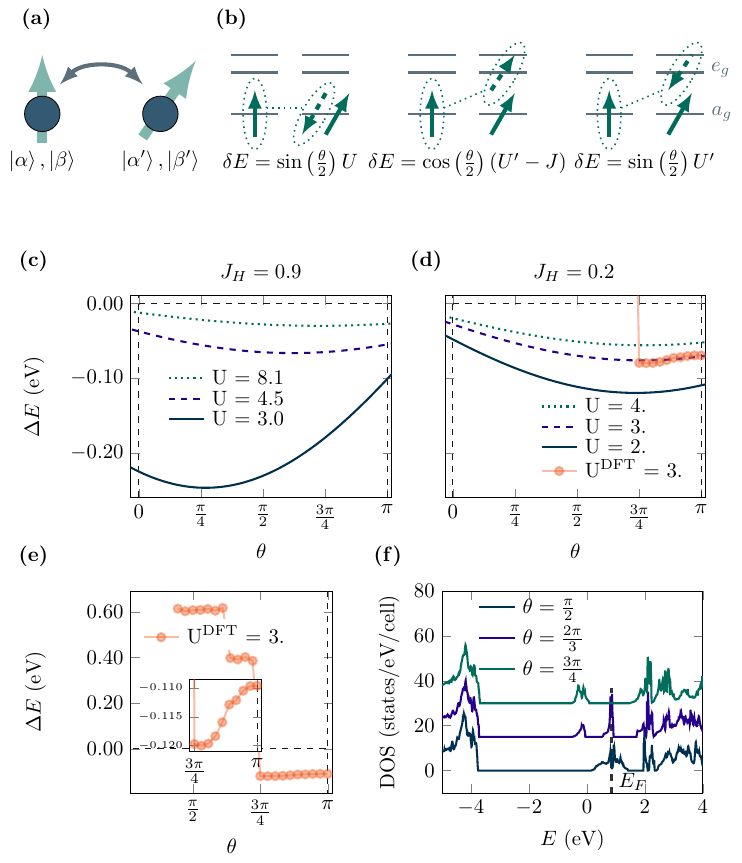}
    \caption{Comparison of a noncollinear magnetic ground state in the double-exchange Hubbard model and density functional theory.
    (a) Electron hopping in the two-site double-exchange Hubbard model with mean-field background spins shown in green. (b) Three different virtual hopping processes and their energy increase starting from an occupation with a single electron in the $a_g$ orbital with spin state $\alpha_i$ on each of the two sites.    
    (c) and (d) total energy versus the angle $\theta$ between the magnetization directions of the two Ti atoms in the primitive unit cell. (e) The angle versus energy plot for the DFT calculations shows jumps at various critical angles which can be resolved from Lifshitz transitions in the density of states, shown in (f). }
    \label{energy}
\end{figure}

The magnetism in TiF$_3$ caused by the unpaired $d$ electrons at Ti ions, is strongly influenced by the strong electronic correlations present in the material. First and foremost, this is a result of the narrow width of the $t_{2g}$ band, where the suppressed kinetic energy fosters the dominance of interaction effects. In the framework of perturbation theory applied to the Hubbard model, Perebeinos and Vogt argued that a transition between ferromagnetic and antiferromagnetic ordering takes place at sufficiently high Hubbard interaction \cite{Perebeinos2004}. However, here we argue that the favored magnetic ordering might instead be non-collinear. To support this idea, we use the double-exchange model \cite{nolting2009quantum} described by the following Hamiltonian, in the absence of interactions
\begin{equation}
    \op{H}_0 = \sum_{ij} \sum_{ab} \sum_\gamma t_{ij}^{ab} \op{c}^\dagger_{i a \gamma} \op{c}_{jb\gamma} - J \sum_i \vec{S}_i \op{\vec{s}}_i.
\end{equation}
The first term describes a conventional electron hopping between sites $i$ and $j$, orbitals $a$ and $b$ and spin states $\gamma$. The second term is the double-exchange term, describing the local interaction of the mean-field spin of the Ti atom $\vec{S}_i$ with the spin $\op{\vec{s}}_i$ of a quasi-free electron. Hence, electronic spins tend to align with the local spin, leading to the absence of a good global quantization axis. In a two-site model, we can construct $\vec{S}_1$ to point along the $z$-axis, while $\vec{S}_2$ is rotated by an angle $\theta$. Instead of up and down spins, we introduce $\left|\alpha_i\right>$ and $\left|\beta_i\right>$ given by
\begin{align}
  \left|\alpha_1\right> &= \left(\begin{array}{c}
       1  \\
       0 
  \end{array}\right),\qquad
  \left|\beta_1\right> = \left(\begin{array}{c}
       0  \\
       1 
  \end{array}\right),\notag \\
    \left|\alpha_2\right> &= \left(\begin{array}{c}
       \cos\left(\frac{\theta}{2}\right) \\
       \sin\left(\frac{\theta}{2}\right) 
  \end{array}\right),\qquad
  \left|\beta_2\right> = \left(\begin{array}{c}
       \sin\left(\frac{\theta}{2}\right) \\
       -\cos\left(\frac{\theta}{2}\right) 
  \end{array}\right).
  \label{spins}
\end{align}
The hopping between the two sites is shown in Figure~\ref{energy}(a). In contrast to the collinear case, there is always a finite hopping amplitude between both sites, regardless of the spin-state. 
The electron-electron interactions are described by the Hubbard model,
\begin{multline}
    \op{H} = \op{H}_0 + \frac{U}{2} \sum_i \sum_a \op{n}_{i a \alpha_i} \op{n}_{i a \beta_i} + \frac{U'}{2} \sum_i \sum_b \sum_{a\neq b} \op{n}_{i a \alpha_i} \op{n}_{i b \beta_i} \\ -   \frac{J_H}{2} \sum_i \sum_b \sum_{a\neq b} \sum_\gamma \op{n}_{i a \gamma} \op{n}_{i b \gamma_i}.
    \label{HubbardHam}
\end{multline}
Besides the non-interacting double exchange Hamiltonian $\op{H}_0$, the Hubbard Hamiltonian contains three terms. The first term describes the repulsion of electrons occupying the same site and same orbital with opposite spins by an energy $U$. The second term describes the repulsion of electrons occupying different orbitals at the same site. The third term, the Hund's coupling, is an attractive term supporting the occupation of different orbitals with the same spin states on the same site. Equation \eqref{HubbardHam} cannot be solved directly, but is discussed using perturbation theory. We assume a ground state with a single electron on each Ti atom, occupying the $a_g$ orbital and spin state $\left|\alpha_i\right>$. Similar to Ref.\cite{Perebeinos2004}, we only consider $(dd\pi)$ hopping with a small hopping strength $t \approx 0.225$~eV. From the orbital overlap \eqref{orbitals1} and spin overlap \eqref{spins}, the hopping amplitude for a process $\left|a_g \alpha\right> \rightarrow \left|a_g \alpha\right>$ between two neighboring sites is $\frac{2}{3} t \sin\left(\frac{\theta}{2}\right)$. The energy increase due to Hubbard interaction is $\delta E = \sin\left(\frac{\theta}{2}\right) U$. In second order perturbation theory, this leads to an energy lowering of $\Delta E = - \sum_{i=1}^6 \frac{\left|\left< a_g \alpha\right|\op{H}_0 \left|a_g \alpha_i\right>\right|^2}{\delta E} = - \frac{8 \sin\left(\frac{\theta}{2}\right) t^2}{3 U}$. Adding the contributions for the processes $\left|a_g \alpha\right> \rightarrow \left|e_g; i\, \alpha\right>$ and $\left|a_g \alpha\right> \rightarrow \left|e_g;\, i \beta\right>$, the total energy lowering is given by
\begin{equation}
    \Delta E = - \frac{8}{3}\frac{ \sin\left(\frac{\theta}{2}\right) t^2}{U} - \frac{4}{3}\frac{ \cos\left(\frac{\theta}{2}\right) t^2}{U'-J} - \frac{4}{3}\frac{ \sin\left(\frac{\theta}{2}\right) t^2}{U'}.
    \label{totalenergy}
\end{equation}
In the limit of ferromagnetism ($\theta = 0$) and antiferromagnetism ($\theta = \pi$), equation \eqref{totalenergy} gives
\begin{align}
    \Delta E^{\text{FM}} &= - \frac{4}{3}\frac{ t^2}{U'-J_H}, \label{EFM}\\
    \Delta E^{\text{AFM}} &= - \frac{8}{3}\frac{ t^2}{U} - \frac{4}{3}\frac{ t^2}{U'}, \label{EAFM}
\end{align}
which is in agreement with the previously derived expressions in Ref. \cite{Perebeinos2004}. Note that the following relationship holds in the atomic limit, $U = U' + 2 J_H$.

The total energy \eqref{totalenergy} evaluated for various values of the Hund's coupling $J_H$ and Hubbard repulsion $U$ is shown in Figure~\ref{energy}. The value $J_H = 0.9$~eV corresponds to the value obtained in Ref.~\cite{Perebeinos2004}. From \eqref{EFM} and \eqref{EAFM} it can be determined that the transition between ferromagnetic and antiferromagnetic ground state takes place at $U/J \approx 4$. Indeed, we can see that this behavior is reflected in our results for the limiting cases of $\theta = 0, \pi$. However, it becomes apparent that the lowest energy configuration for each Ti-Ti pair is a noncollinear order, with an optimal magnetization angle depending on the microscopic parameters $U$ and $J_H$. The asymptotic value of $\theta$ in the limit $U \rightarrow \infty$ is $\theta_{U\rightarrow \infty} = 2 \arctan 3 \approx \frac{4}{5} \pi$.  The non-collinear order results from the multi-orbital nature of TiF$_3$. In the absence of the $e_g$ orbital, only the first term in the total energy \eqref{totalenergy} remains. This term is minimized for the antiferromagnetic order ($\theta = \pi$), which corresponds to a Mott insulator at half-filling.

We verify this behavior by performing non-collinear density functional theory calculations for the primitive lattice with 2 Ti atoms, using the Vienna Ab Initio Simulation Package (VASP) \cite{a22}. The exchange-correlation functional was approximated by the generalized gradient approximation (GGA) \cite{a24}. We used a $10\times10\times8$ $\vec{k}$-point mesh.
We computed the total energy for a constrained magnetization angle varying between $100^\circ$ and $180^\circ$ degrees, \change{introducing a Lagrange multiplyer of $\lambda = 0.2~\text{eV}/\mu_B$ \cite{Ma2015}. Furthermore, we used a} Hubbard-U of 3~eV and a Hund coupling of 0.2~eV. \change{The cut-off energy was 290~eV.} In agreement with our model, the total energy is minimized for a magnetization angle of $\frac{3}{4}\pi$ between the two Ti atoms in the primitive cell, as shown in Figure~\ref{energy}(b) and Figure~\ref{energy}(c).  In contrast to the simplified model, the ab initio calculations show sharp steps in the total energy, as can be seen in Figure~\ref{energy}(c). To explain these steps, we calculated the density of states for three angles, $\theta = \frac{\pi}{2}, \theta=\frac{2}{3}\pi$, and $\theta = \frac{3}{4}\pi$.  While the system remains insulating for angles larger than $\theta = \frac{3}{4}\pi$ ($\approx$ antiferromagnetic) it cuts a band for smaller angles. Interestingly, several such transitions seem to take place, leading to a step-wise increase in the total energy. 

\section{Experimental evidence for insulating behavior}
To examine possible insulating behavior of TiF$_3$, we present frequency-dependent far-infrared reflectivity conducted on a hydraulically-pressed pellet using a combination of beamsplitters and helium-cooled bolometer detectors to cover the spectral range where strong signatures of single-phonon lattice excitations and significant free carrier contributions are expected to be obvious, if present. The clear presence of strong reststrahlen bands shows that vibrations dominate the polarizability of this material. These strong lattice vibrations can be fit to a set of five Lorentz oscillators in this spectral range and are similar to those observed in single crystals of the related compound ScF$_3$\cite{Handunkanda2019}. Kramers-Kronig consistency in guaranteed by the form of the Lorentz oscillator. Importantly, no free carrier contribution is used in the fitting procedure, consistent with a vanishingly small Drude weight expected for an insulating material. We conclude that the infrared response of TiF$_3$ is inconsistent with a scenario of metallic conduction and is consistent with our attempts to measure directly the resistivity of the sample. In addition, an analysis of magnetic susceptibility data appears to show no long-range antiferromagnetic order down to 10 K \cite{Perebeinos2004,sheets2023}.

\begin{figure}
	\includegraphics[width=0.34\textwidth]{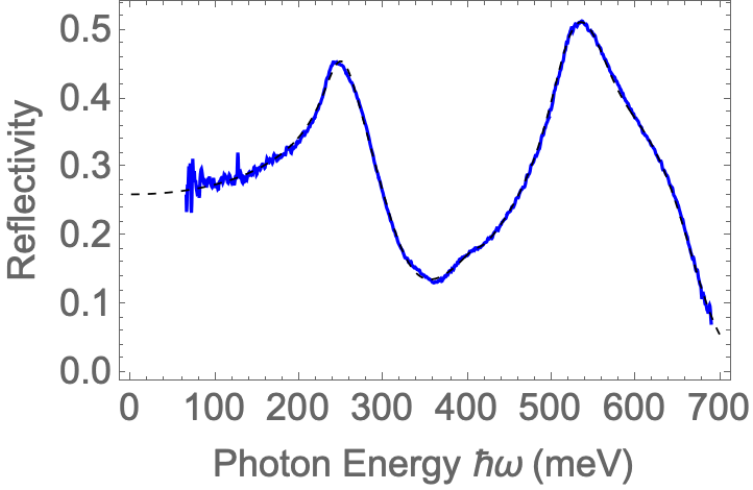}
	\caption{Optical Reflectivity of hydraulically-pressed pellets of TiF$_3$. Two broad reststrahlen bands show that the system is insulating.}
	\label{fig:TiF3Reflectivity}
\end{figure}

\section{First Principles Electronic and Magnetic Structure}

\begin{figure}[b]
    \includegraphics[width=0.45\textwidth]{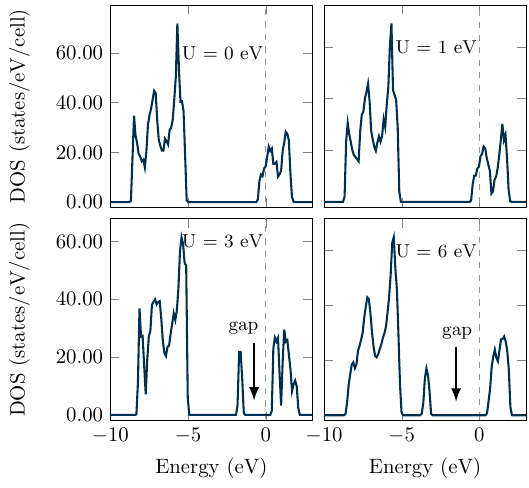}
    \caption{Density of states (DOS) from non-collinear magnetic GGA+U VASP calculations using a unit cell of Ti$_6$F$_{18}$ for selected values of $U$ and $J_H~ 0.0-0.4$. The Fermi level is chosen to be at zero energy for $U \le 1$ and is in the gap above zero energy for higher U. Note the metallic behavior at $U= 0, 1$ and the insulating behavior for $U\ge 3$ eV. }
    \label{dos}
\end{figure}
In order to reveal the metal-insulator transition depending on the interaction strength and non-collinear magnetism, we have performed additional VASP calculations both on a full rhombohedral unit cell (space group R$\overline{3c}$, $\#$167) and on a primitive cell.  In particular, we have used a larger unit cell containing 6 Ti and 18 F atoms, in order to probe possible non-collinear magnetism. 
For the large cell, a $6\times 6\times 6$ $\vec{k}$-space $\Gamma$ centered mesh according to Monkhorst and Pack \cite{a25} was used during the self-consistent cycle. 
Structural optimization was performed until the Hellman-Feynman forces acting on the atoms were negligible.

The density of states plots in Fig.{\ref{dos}} illustrate a metal-insulator transition taking place as a function of the Hubbard $U$ parameter in the extended cell. At higher $U$ values, the insulating gap is seen to increase due to increasing $d$-orbital repulsion. As we see here, a large $U$ value such as $U=8.1$~eV used in
Ref.~\cite{Perebeinos2004}, is not necessary to yield an insulating state. This value is also significantly higher than the Hubbard repulsion $U=3.3$~eV used by Mattsson and Paulus \cite{mattsson2019density}. The latter value is more consistent with typical values of Hubbard repulsion $U \approx 3-4$~eV obtained for various transition metal trihalides \cite{Yekta2021,esteras2021hubbard,he2016unusual}. We note that recent frequency dependent far-infrared reflectivity measurements support an insulating state in TiF$_3$ \cite{sheets2023}.

First principles-based non-collinear magnetic calculations of TiF$_3$ extend the capability of revealing the interplay between electronic and magnetic properties. Typically, non-collinear structures are found in materials where the topology/geometry of the arrangement excludes collinear order \cite{sandratskii1996noncollinear,Sandratskii1998,bousquet2016non,zlotnikov2021aspects}. This could be possibly due to frustrations arising from a competition of ferromagnetic and antiferromagnetic interactions, for example, as in triangular lattices. 
Previous computational work discussing the electronic and magnetic structure of TiF$_3$ confined the magnetism to collinear magnetic order \cite{Perebeinos2004,mattsson2019density}. In agreement with Ref.~\cite{Perebeinos2004}, we find that for a collinear spin arrangement with structural optimization, Ti$_2$F$_6$ VASP calculations predict metallic, ferromagnetic behavior at $U=1$~eV, $J_H=0.2$~eV while at $U=3$~eV or higher, with similar $J_H$ values, an insulating state is seen from the density of states. The magnetic moment in the ferromagnetic phase is about 1 $\mu_B$ per Ti atom.

\begin{figure}
    \centering
    \includegraphics[width=0.49\textwidth]{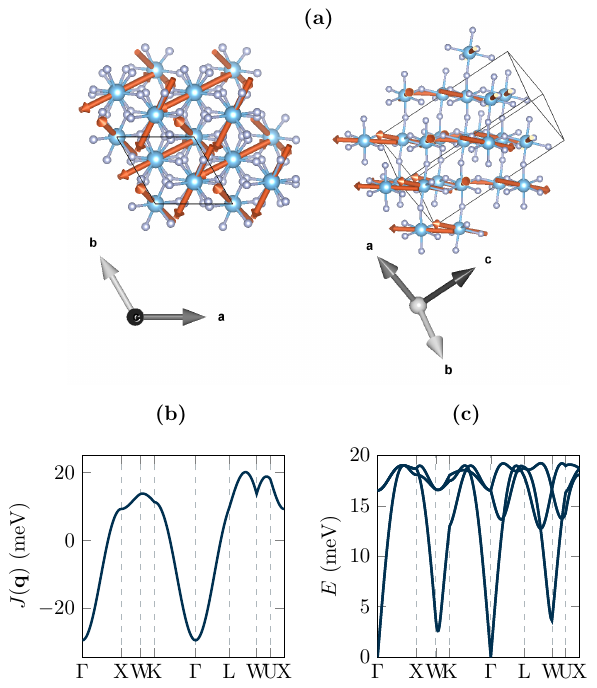}
    \caption{
    (a) Magnetic structure at $U=2$ eV for Ti$_6$F$_{18}$ from non-collinear magnetic calculations using VASP. These moments (shown in red) lie in a plane for $U=2$ eV while at larger $U$ values some of the moments point out of this plane indicating possible magnetic frustration. Magnetic order seen here depends on the initial choice of the magnetic orientations used.
    (b) Fourier transform of the Heisenberg exchange parameters ($U=3.3$~eV). The maximum between the $L$ and $W$ points in the Brillouin zone points towards an incommensurate magnetic order or frustration. \change{(c) Magnon spectrum computed for an antiferromagnetic ground state.} }
    \label{fig:Jq}
\end{figure}

When a larger unit cell (Ti$_6$F$_{18}$) is used to carry out first principles-based VASP GGA+U calculations, non-collinear magnetism was found. 
In Ti, the spin-orbit interaction is not large and of the order of 20-30 meV or less. However, this is most likely sufficient to introduce changes in magnetic order at low temperature. With small $U$ values (of the order of $2$~eV), a metallic state is seen with some magnetic order which is neither completely ferromagnetic nor antiferromagnetic. These local moments appear to lie on a certain crystallographic plane (as seen in Fig. ~\ref{fig:Jq}(a) - $(1{\bar 1}$1) plane) with pairs of Ti atoms antiferromagnetically aligned at different angles with respect to each pair but on a planar configuration. However, there could be some sensitivity to the chosen initial magnetic configuration here. At larger $U$ values ($U \ge 3$~eV), the system is insulating with some of these moments pointing out of the above-mentioned plane, indicating possible magnetic canting and frustration.

To estimate the magnetic ground structure, we have calculated Heisenberg exchange coupling constants $J_{ij}$ using  the magnetic force theorem implemented within a first-principles Green function method~\cite{Liechtenstein1987,Hoffmann2020}. \change{We note that this method is complementary to our VASP calculations and has been consitently shown to give similar results in electronic structure calculations \cite{Hoffmann2020}.} The full Fourier transform of the exchange constants, $J(\vec{q})$, is presented in Fig.~\ref{fig:Jq}(b).  $J(\vec{q})$ is proportional to the transversal magnetic susceptibility and its maximum corresponds to the total energy minimum. A maximum of $J(\vec{q})$ at a high symmetry point would reflect a commensurate magnetic structure of ferro- (if the maximum is at $\Gamma$ point) or an antiferromagnetic order. In the case of TiF$_3$ the maximum of $J(\vec{q})$ was found between $L=(1/2,1/2,1/2)$ and $W=(1,1/2,0)$ high symmetry points, which corresponds to a noncollinear magnetic order with the ordering vector $\vec{Q}=(1,0.32,0.18)$. The origin of the noncollinear structure is the short-range nature of the magnetic interaction. For the case of $U=3.3~\text{eV}$, the exchange interaction between the nearest magnetic moments is about -0.78 meV, while other coupling constants were found to be negligibly smaller \change{(magnitude $< 0.02~\text{meV}$)}. Thus, a dominant antiferromagnetic interaction between the nearest moments leads to a frustration of magnetic moments and forming  a noncollinear magnetic order. Our calculations showed that the obtained $J_{ij}$ depends weakly on the choice of the Hubbard parameter $U$ ($2~\text{eV} \leq U \leq 6~\text{eV}$). Thereby, the critical transition temperature was estimated in the range of 18-22~K.

\section{Conclusions}

In agreement with conclusive experimental reflectivity measurements, 
we argue that TiF$_3$ is an insulator, with a gap due to strong electron-electron correlations. Due to the absence of a clear antiferromagnetic signal in the material \cite{sheets2023}, we explored a potential magnetic frustration or the tendency towards non-collinear magnetism. Here we extended the model of Perebeinos and Vogt \cite{Perebeinos2004}, showing that a non-collinear arrangement of magnetic moments between neighboring Ti ions might be favorable. This claim is furthermore corroborated by density functional theory calculations that include explicit correlations. First, we observed a non-collinear magnetic ordering by performing total energy calculations using a pseudopotential method. Furthermore, we computed Heisenberg exchange parameters and found a strong antiferromagnetic nearest neighbor coupling with all other exchange parameters being vanishingly small. The Fourier transform of the Heisenberg exchange clearly points towards an incommensurate magnetic order. Yet, the exact magnetic configuration seems to be highly sensitive to input parameters, leading to open questions such as the possible existence of a spin liquid phase.

R.M.G. acknowledges support from the Swedish Research Council (VR starting Grant No. 2022-03350) and Chalmers University of Technology. J.N.H. acknowledges support from the U.S. National Science Foundation, award No. NSF-DMR-1905862. A.E. acknowledges funding by Fonds zur Förderung der
Wissenschaftlichen Forschung (FWF) Grant No. I 5384. We also acknowledge the computing resources provided by the Center for Functional Nanomaterials, which is a U.S. DOE Office of Science Facility, at Brookhaven National Laboratory under Contract No. DE-SC0012704 and the Swedish National Infrastructure for Computing (SNIC) via the National Supercomputer Centre (NSC). 

\providecommand{\noopsort}[1]{}\providecommand{\singleletter}[1]{#1}%

\end{document}